\documentclass[aps,prl,reprint,superscriptaddress]{revtex4-1}

\usepackage{graphicx}
\usepackage{subfigure}
\usepackage{color}
\usepackage{eqnarray, amsmath}
\usepackage{amssymb}
\usepackage{epstopdf}
\usepackage[linktocpage,colorlinks=true,linkcolor=blue,citecolor=blue,breaklinks=true]{hyperref}
\usepackage{verbatim}
\usepackage{breakcites}
\usepackage[version=3]{mhchem}

\newcommand{\be}{\begin{equation}}
\newcommand{\ee}{\end{equation}}

\begin{document}

\preprint{}

\title{The combined effects of shear and buoyancy on phase boundary stability}

\author{Srikanth Toppaladoddi}
\affiliation{University of Oxford, Oxford OX2 6GG, UK}
\affiliation{Yale University, New Haven, CT 06520, U.S.A.}

\author{J. S. Wettlaufer}
\affiliation{Yale University, New Haven, CT 06520, U.S.A.}
\affiliation{University of Oxford, Oxford OX2 6GG, UK}
\affiliation{Nordita, Royal Institute of Technology and Stockholm University, SE-10691 Stockholm, Sweden}

\email[]{john.wettlaufer@yale.edu}

\date{\today}

\begin{abstract}
We study the effects of externally imposed shear and buoyancy driven flows on the stability of a solid-liquid interface. By reanalyzing the data of Gilpin \emph{et al.} [\emph{J. Fluid Mech.}, {\bf 99}(3), 619 (1980)] we show that the instability of the ice-water interface observed in their experiments was affected by buoyancy effects, and that their velocity measurements are more accurately described by Monin-Obukhov theory.  A linear stability analysis of shear and buoyancy driven flow of melt over its solid phase shows that buoyancy is the only destabilizing factor and that the regime of shear flow here, by inhibiting vertical motions and hence the upward heat flux, stabilizes the system. It is also shown that all perturbations to the solid-liquid interface decay at a very modest strength of the shear flow.  However,  at much larger shear, where flow instabilities coupled with buoyancy might enhance vertical motions, a re-entrant instability may arise.
\end{abstract}

\pacs{}

\maketitle

\section{Introduction}

Flow of a melt over its solid phase can profoundly influence the latter's evolution and stability \citep[e.g.,][]{epstein1983, glicksman1986, davis1990, huppert1990, worster2000}. Examples  abound in both natural \citep[e.g.,][]{Untersteiner:1965, wettlaufer1991, mcphee_book, meakin2009, solari2013, ramudu2016, claudin2017} and engineering \citep[e.g.,][]{delves1968, delves1971, forth1989} settings. Flows over phase-changing boundaries can be grouped into the following two categories: (1) free flows, which arise due to density differences created during solidification \citep{davis1984, liu1996, wettlaufer1997, worster1997, wykes2018}, and (2) forced flows, which are typically shear driven, and are introduced to control morphological and/or hydrodynamical instabilities \citep{delves1968, delves1971, coriell1984, forth1989}. 

In the absence of an external flow, the rates of freezing are typically sufficiently large so that a planar solid-binary liquid interface will become highly convoluted, leading to one of the two components being trapped in the interstices of the crystals of the other component \citep{worster2000}.  In engineering the imposition of a flow was motivated by controlling the instability, whereas in natural settings it is often an unavoidable part of the environment \citep[e.g.,][]{delves1968, delves1971, coriell1984, forth1989, schulze1994, schulze1995, schulze1996, feltham1999, feltham2002, neufeld2006, neufeld2008, neufeld2008shear, camporeale2012}.  Here, we focus on understanding the effects of shear and buoyancy on directional solidification of a pure melt. However, we shall review the results on directional solidification of binary mixtures as well as those for pure melts, because there are some commonalities in the dynamics of the two systems.

%
Some of the first studies to investigate the effects of shear-driven flows on directional solidification of a binary alloy using linear stability analysis are those of \citet{delves1968, delves1971} and \citet{coriell1984}. \citet{delves1968, delves1971} studied the effects of a parabolic flow on morphological instability and found that the flow suppresses the instability, with the degree of suppression depending on the material considered. He also found that the flow gives rise to travelling waves along the interface. \citet{coriell1984} studied the effects of Couette flow on the morphological and thermosolutal instabilities during directional solidification of a lead-tin alloy. Their findings suggest that Couette flow suppresses the onset of thermosolutal instability to a larger degree than the onset of morphological instability. However, the use of Couette flow as the base-state velocity profile seems incompatible with the momentum-balance equations \citep{coriell1980}, which admit the asymptotic suction boundary-layer profile \citep{drazin2004} as their solution.

\citet{forth1989}
studied directional solidification of a binary alloy in the presence of an asymptotic-suction-boundary-layer flow. They focused on (a) understanding how the fluid flow affects the morphological instability, and (b) understanding how the freezing interface affects the shear flow instability. Under certain conditions they find that the shear flow only leads to the generation of traveling waves along the interface, and that the speed of these waves varies linearly with the imposed flow speed. However, under the same conditions, the freezing interface was found to have negligible effects on the hydrodynamic instability. 

The structure resulting from the instability of the solid-binary liquid interface is known as a mushy layer \citep{worster2000}, and is modeled as a chemically reacting porous medium \citep{worster1991}. The most common example of mushy layer is the sea ice found in Earth's polar regions \citep{feltham2006}. Here, compositional convection can be induced both in the mushy layer, which contains brine trapped between ice crystals, and in the sea water, which is gravitationally unstable due to high concentration of salt -- rejected during solidification -- close to the ice-water interface \citep{worster1992}. These modes of convection are termed mushy and boundary layer modes, respectively \citep{worster1992}, and have been observed in the laboratory \citep{wettlaufer1997}. It is intuitive that in the presence of a shear flow, the evolution of any incipient perturbation at the mush-liquid interface should depend on the interaction between the flows in the melt and mushy layer.

By neglecting the effects of buoyancy in both the bulk melt and the mushy layer, \citet{feltham1999} investigated the effects of forced flow of inviscid and viscous melts on the morphology of a mushy layer. They found that an external flow over a corrugated mush-liquid interface results in a pressure perturbation along the interface that drives flow in the mushy layer, and under certain conditions this leads to the growth of the perturbations with a wavelength commensurate with the depth of the mush layer. The perturbed heat flux from the liquid was found to have no influence on the evolution of the perturbation and was only responsible for introducing traveling waves at the interface.

\citet{neufeld2008, neufeld2008shear} studied the effects of shear flow on the mushy- and boundary-layer modes of convection using both theory and experiments. They found that; (1) Below a critical value of the shear-flow velocity, both modes of convection are moderately suppressed; (2) Above a critical shear-flow velocity, the stability of both modes of convection decreases monotonically with the strength of the flow; (3) For sufficiently strong shear flow, striations of zero solid fraction transverse to the flow direction are generated. These striations are quasi-two-dimensional and form because of localized dissolution and growth of the mushy layer, which in turn is due to the interplay between shear and buoyancy.

Relative to binary mixtures, there have been far fewer studies of the influence of external flows on the directional solidification of pure melts. One of the first experimental studies was by \citet{Gilpin1980}, who investigated the evolution of a layer of pure ice in contact with a turbulent flow  in a closed-loop water tunnel with an upper free surface. A layer of ice rests over a surface that is maintained at a temperature less than the melting temperature and a shear flow is maintained over the ice layer, with the far-field temperature greater than the melting point. Before starting the flow, the ice-water interface was perturbed by melting a groove into the ice layer. Under certain conditions, the perturbation at the ice-water interface was observed to grow, leading to the formation of a ``rippled" surface. They found that the heat transfer rate over the rippled surface was $30$--$60 \%$ larger than that on a planar surface and the evolution of the ice layer was wholly attributed to the overlying shear flow. \citet{Gilpin1980} also performed a linear stability analysis of their system to explain the observed instability. However, instead of solving for the stability equations in the fluid region, they represented the effects of the flow using a perturbed ``heat-transfer coefficient", whose amplitude and phase were obtained by fits to experimental data. This approach would be difficult to justify as fluctuations in a turbulent flow cannot be assumed to be small. However, one crucial point that \citet{Gilpin1980} evidently overlooked is that because the far-field temperature of water was greater than the melting point, the water column above the ice layer was unstably stratified due to the $4$ $^{\circ}$C density maximum, which can exert a controlling influence on heat flux \citep{veronis1963, TW2018}.   

Here, motivated by the experiments of \citet{Gilpin1980}, we study the effects of shear and buoyancy on the phase evolution of a pure melt. Specifically, we study solidification of a pure melt in the presence of Couette flow and Rayleigh-B\'enard convection. The reason for our choice of the Rayleigh-B\'enard-Couette system is two-fold: (1) To have an analytically tractable system where the relative effects of shear and buoyancy on the stability of the phase boundary can be studied; and (2) To ascertain whether the instability observed by \citet{Gilpin1980} can indeed be found in the linear regime of such a system. Because the velocity profile in the viscous sublayer varies linearly with the distance from the wall \citep{Monin1971} the problem we study has the key features of that in \citet{Gilpin1980}. We also show that the velocity measurements of \citet{Gilpin1980} are better explained by Monin-Obukhov theory \citep{Monin1971}, which describes turbulent shear flow in stratified fluids. We then perform a linear stability analysis and study the effects of shear and buoyancy on the growth of perturbations at the solid-liquid interface.

\section{Re-analysis of the experimental results of Gilpin \emph{et al.} \cite{Gilpin1980}}
In this section, we use the same notation as did \citet{Gilpin1980} to describe their results.
\subsection{Details of the experiments}
\begin{figure}
\begin{centering}
\includegraphics[trim = 150 100 150 150, width = 0.65\linewidth]{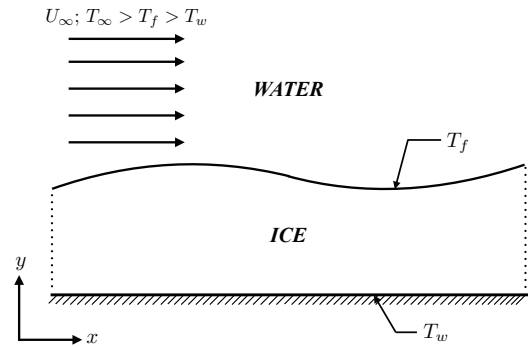} 
\caption{Schematic of the experimental study of \cite{Gilpin1980}.}
\label{fig:gilpin_schematic}
\end{centering}
\end{figure}

Figure \ref{fig:gilpin_schematic} shows a schematic of the experimental study of \citet{Gilpin1980}. The bottom wall is maintained at a temperature $T_w$, the ice-water interface is at the bulk equilibrium temperature $T_f$, and the far-field temperature $T_{\infty}$, is such that $T_{\infty} > T_f > T_w$. The far-field flow speed is $U_{\infty}$.
Because pure water has a density maximum at $4$ $^{\circ}$C, the water column above the ice layer is unstably stratified.

\subsection{Monin-Obukhov (M-O) theory for a smooth surface}
In wall-bounded turbulent shear flows of neutrally buoyant fluids, the flow consists of the inner and the outer regions \citep{Monin1971, sreenivasan1989}. The inner region is subdivided into: (1) the viscous sublayer, which is closest to the wall, where the effects of viscosity are dominant; (2) the buffer layer, which is next to the viscous sublayer, where viscous and inertial effects are equally important; and (3) the log-layer, where neither the effects of the wall nor that of the outer region are important. In the limit of asymptotically large Reynolds number, scaling arguments for the behavior of the mean horizontal velocity, $U$, in the log-layer lead to \citep{Monin1971, sreenivasan1989}:
\be
U^+(y^+) = \frac{1}{k_s} \, \log(y^+) + B,
\label{eqnlog}
\ee
where $k_s = 0.41$ is the K\'arm\'an constant, and $B = 5.5$ is another constant. The constants $k_s$ and $B$ are believed to be universal, but their values have been determined only empirically \citep{sreenivasan1989}. The superscript $^+$ denotes non-dimensionalization by $u_*$, the friction velocity, and $l_v = \nu/u_*$, the viscous length scale.

In the case of wall-bounded shear flows of stratified fluids, the stratification affects the mean velocity as follows. If the flow is unstably stratified, there are more vigorous vertical motions and thus more vertical mixing. Hence, the mean velocity at any location is smaller than that for a neutrally buoyant fluid at the same location. However, if the flow is stably stratified then vertical motions are suppressed, leading to a mean velocity that is larger than that for a neutrally buoyant fluid \citep[e.g.,][]{Monin1971, turner1979}. 

\begin{figure}
\begin{centering}
\includegraphics[width = \linewidth]{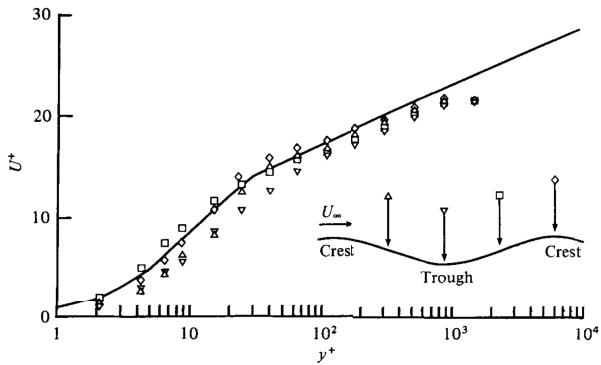} 
\caption{Velocity measurements from Fig. 4 of \citet{Gilpin1980} for $Re_{\delta} = 11000$, where $Re_{\delta}$ is the Reynolds number based on the thickness of the boundary layer. From left to right, the positions in the inset are numbered 1, 2, 3, and 4. The solid line is the Law of the Wall \citep{Monin1971}.}
\label{gilpin}
\end{centering}
\end{figure}

Figure \ref{gilpin} shows $U^+(y^+)$ at different locations in the experiments of \citet{Gilpin1980}; the mean velocity profiles show a systematic deviation from the log-layer, indicating unstable stratification.
Moreover, the amplitude of the rippled interface they observed was small compared to its wavelength. Hence, we treat the surface as planar for the purpose of quantifying the effects of stratification, for which we extend the M-O theory.

The relative effects of inertia and buoyancy are represented by the M-O length scale, denoted by $\mathcal{L}$ \citep{Monin1971}. For stable stratification $\mathcal{L}>0$ and for unstable stratification $\mathcal{L}<0$, with the effects of stratification being important for distances $y > O(|\mathcal{L}|)$ from the wall. 
Following \citet{Monin1971}, we let $\xi = y/\mathcal{L}$ and write
\be
\frac{\partial U}{\partial y} = \frac{u_*}{k_s \,\mathcal{L}} \, f\left(\xi\right) \equiv  \frac{u_*}{k_s \, y} \, \phi(\xi),  
\label{eqn1}
\ee
where $f$ is an unknown function of $\xi$ and $\phi(\xi) = \xi f(\xi)$.  
Scaling equation \ref{eqn1} with $u_*$ and $l_v$, we have:
\be
\frac{\partial U^+}{\partial y^+} = \frac{1}{k_s \, y^+} \, \phi\left(\frac{y^+}{\mathcal{L}^+}\right).
\label{eqn3}
\ee
For $y^+/\mathcal{L}^+ \ll 1$, $\phi$ can be expanded in a power series: $\phi = 1 + \beta \frac{y^+}{\mathcal{L}^+} + \textrm{h.o.t.}$ Using this in Eq. \ref{eqn3} and integrating with respect to $y^+$ gives
\be
U^+ = \frac{1}{k_s} \log(y^+) + \beta \frac{y^+}{k_s \, \mathcal{L}^+} + A.
\label{eqn4}
\ee
In the limit $\mathcal{L} \rightarrow \infty$, equation \ref{eqn4} should reduce to the classical law of the wall, which gives $A = B = 5.5$. As this analysis is valid for distances `far away' from the wall, the value of $\beta$ is taken to be $0.6$ \citep{Monin1971}.
%

\subsection{Comparison with the experiments}
\begin{figure}
\begin{centering}
\includegraphics[trim = 0 0 0 0, width = \linewidth]{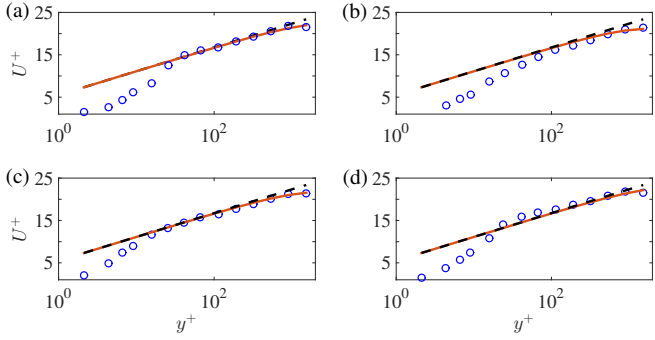} 
\caption{Comparison of the theory (equation \ref{eqn4}) with the measurements of \citet{Gilpin1980} at (a) Position 1, (b) Position 2, (c) Position 3, and (d) Position 4. Circles: Data from \citet{Gilpin1980}; dashed line: $U^+ = \frac{1}{k_s} \log(y^+) + A$; solid line: $U^+ = \frac{1}{k_s} \log(y^+) + b \,  y^+ + A$. Here, $b = -0.00092$, $-0.0015$, $-0.0012$ and $-0.0008$ at positions 1, 2, 3, and 4, respectively.}
\label{fig:gilpin1}
\end{centering}
\end{figure}

The velocity profile given by equation \ref{eqn4} can now be fit to the data of \citet{Gilpin1980}. Here, $b = \beta/(k_s \, \mathcal{L}^+)$ is the only fitting parameter, with $\beta$ and $k_s$ already known. Figure \ref{fig:gilpin1} shows the fits of equation \ref{eqn4} to the data in figure \ref{gilpin}.
The averaged value of $b$ from the fits to the data at the four positions is $b_{\tiny\textrm{avg}} = - 0.0011$, and hence $\mathcal{L}^+_{\tiny\textrm{avg}} = \beta/(k_s \, b) = -1330.38$. Thus, because $\mathcal{L}_{\tiny\textrm{avg}} < 0$, we confirm that the water column was unstably stratified. From the range of values given for the free-stream velocity in the \citet{Gilpin1980} experiments, we take $U_{\infty} = 0.5$ ms$^{-1}$ and use their equation 13, 
\be
u_*/U_{\infty} = 0.229 \, Re_{\delta}^{-0.132},
\ee
to obtain $u_* = 0.033$ ms$^{-1}$ for $Re_{\delta} = 11000$. Taking $\nu = 10^{-6}$ m$^2$s$^{-1}$, we obtain $l_v = 29.93$ $\mu$m, and hence $|\mathcal{L}_{\tiny\textrm{avg}}| = |\mathcal{L}^+_{\tiny\textrm{avg}}| \, l_v = 0.04$ m. The height of the test section reported is $0.457$ m, which makes $|\mathcal{L}_{\tiny\textrm{avg}}|$ about $9$\% of the test-section height. However, because of the departure of velocity profiles from the classical log-law at smaller distances from the ice surface, the value of $|\mathcal{L}_{\tiny\textrm{avg}}|$ estimated here may be larger than the actual value.
 
\section{Governing Equations}
\begin{figure}
\begin{centering}
\includegraphics[trim = 10 100 10 0, width = \linewidth]{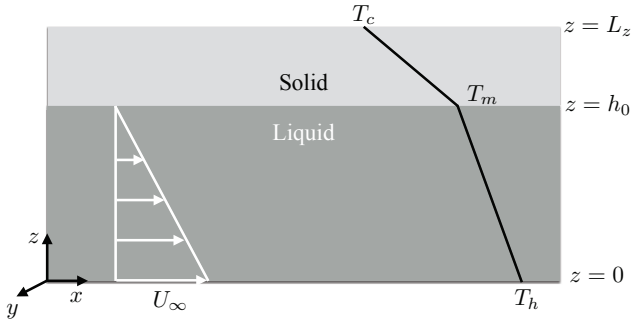} 
\caption{Schematic of the domain considered here.}
\label{fig:stefan2}
\end{centering}
\end{figure}

To perform a linear stability analysis, we consider the domain shown in figure \ref{fig:stefan2}. The length of the cell is $L_x$ and the depth of the cell is $L_z$. At the initial instant the solid occupies the region $h_0 \le z \le L_z$, and the liquid occupies $0 \le z \le h_0$. The solid-liquid interface is planar and is at $z = h_0$. The initial thickness of the solid layer is $d_0$, and hence $L_z = h_0 + d_0$. The upper surface is maintained at a temperature $T_c$ and the lower surface is maintained at $T_h$. The temperatures are such that $T_h > T_m > T_c$, where $T_m$ is the melting temperature of the solid. The liquid considered has a linear equation of state, hence the liquid column in unstably stratified. The bottom surface moves at a constant horizontal velocity $U_{\infty}$, as shown in figure \ref{fig:stefan2}.
The governing equations in the different regions are as follows.

\subsection{Liquid}
The continuity, Boussinesq, and heat-balance equations are
\be
\nabla \cdot \boldsymbol{u} = 0,
\label{eqn:mass}
\ee
\be
\frac{\partial \boldsymbol{u}}{\partial t} +  \boldsymbol{u} \cdot \nabla  \boldsymbol{u} = -\frac{1}{\rho_0} \, \nabla p + g \, \alpha \, \left(T_l-T_m\right) \, \boldsymbol{k} + \nu \, \nabla^2  \boldsymbol{u},
\label{eqn:NS_liquid}
\ee
\be
\frac{\partial T_l}{\partial t} +  \boldsymbol{u} \cdot \nabla  T_l = \kappa \, \nabla^2  T_l,
\label{eqn:heat_liquid}
\ee
respectively. Here, $\boldsymbol{u}(\boldsymbol{x},t) = (u, v, w)$ is the velocity field, $\rho_0$ is the reference density, $p(\boldsymbol{x},t)$ is the pressure field, $g$ is acceleration due to gravity, $\alpha$ is the thermal expansion coefficient, $T_l(\boldsymbol{x},t)$ is the temperature field, $\nu$ is the kinematic viscosity, and $\kappa$ is the thermal diffusivity.  To simplify matters, we assume the liquid and solid phases have the same density ($\rho_0$) and thermal diffusivity ($\kappa$).

\subsection{Solid}
The temperature field in the solid, $T_s(\boldsymbol{x},t)$, is governed by diffusion viz., 
\be
\frac{\partial T_s}{\partial t} = \kappa \, \nabla^2  T_s, 
\label{eqn:heat_solid}
\ee

\subsection{Solid-liquid interface}
At the solid-liquid interface, we have the Stefan condition
\be
\rho_0 \, L_s \, \frac{\partial h}{\partial t} =  \boldsymbol{n} \cdot \left[\boldsymbol{q_s} - \boldsymbol{q_l}\right]_{z = h_0},
\label{eqn:stefan_full}
\ee
where $L_s$ is the latent heat of fusion, $\boldsymbol{n}$ is the unit vector pointing into the liquid, $\boldsymbol{q_s} = -k \, \nabla T_s\vert_{z=h^+}$ is the heat flux away from the interface into the solid and $\boldsymbol{q_l} = -k \, \nabla T_l\vert_{z=h^-}$ is the heat flux towards the interface from the liquid.

\subsection{Boundary conditions}
The boundary conditions for heat equation in the solid are
\be
T_s(z = L_z, t) = T_c \quad \mbox{and} \quad T_s(z = h_0, t) = T_m, 
\ee
and those for the advection-diffusion equation in the liquid are
\be
T_l(z = 0, t) = T_h \quad \mbox{and} \quad T_l(z = h_0, t) = T_m.
\ee
The velocity field satisfies
\be
u(z = 0, t) = U_{\infty}; \, v(z = 0, t) = w(z = 0, t) = 0,
\ee
and
\be
u(z = h_0, t) = v(z = h_0, t) = w(z = h_0, t) = 0.
\ee

We non-dimensionalize these equations by choosing $U_{\infty}$ as the velocity scale; $h_0$ as the length scale, $t_0 = h_0^2/\kappa$ as the time scale, $p_0 = \rho_0 \, U_{\infty} \, \kappa/h_0$ as the pressure scale, and $\Delta T = T_h - T_m$ as the temperature scale. Using these in equations \ref{eqn:NS_liquid}, \ref{eqn:heat_liquid}, \ref{eqn:heat_solid} and \ref{eqn:stefan_full}, and maintaining the pre-scaled notation, we have
\be
\nabla \cdot \boldsymbol{u} = 0;
\label{eqn:mass_scaled}
\ee
\be
\frac{\partial \boldsymbol{u}}{\partial t} +  Pe \, \left(\boldsymbol{u} \cdot \nabla  \boldsymbol{u}\right) = -\nabla p + \frac{Ra \, Pr}{Pe} \, \theta_l \, \boldsymbol{k} + Pr \, \nabla^2  \boldsymbol{u};
\label{eqn:NS_liquid_scaled}
\ee
\be
\frac{\partial \theta_l}{\partial t} +  Pe \, \left(\boldsymbol{u} \cdot \nabla  \theta_l \right) = \nabla^2  \theta_l;
\label{eqn:heat_liquid_scaled}
\ee
\be
\frac{\partial \theta_s}{\partial t} = \nabla^2  \theta_s;
\label{eqn:heat_solid_scaled}
\ee
and
\be
\frac{\partial h}{\partial t} =  \frac{1}{\Lambda \, \mathcal{S}} \, \left[\boldsymbol{n} \cdot \left(\boldsymbol{q_s} - \boldsymbol{q_l}\right)\right]_{z = 1}, 
\label{eqn:stefan_scaled}
\ee
where,
\be
\theta_l = \frac{T_l - T_m}{\Delta T} \quad \mbox{and} \quad \theta_s = \frac{T_s - T_m}{\Delta T}.
\ee
There are five governing parameters, which are 
\be
Ra = \frac{g \, \alpha \, \Delta T \, h_0^3}{\nu \, \kappa}, \quad Pe = \frac{U_{\infty} \, h_0}{\kappa}, \quad Pr = \frac{\nu}{\kappa},
\ee
\be
\mathcal{S} = \frac{L_s}{C_p \, \left(T_m-T_c\right)} \quad \mbox{and} \quad \Lambda = \frac{\left(T_m - T_c\right)}{\Delta T}.
\ee
where, $Ra$, $Pe$, $Pr$, and $\mathcal{S}$ are the Rayleigh, P\'eclet, Prandtl, and Stefan numbers, respectively. The ratio of the temperature differences across the liquid and the solid regions is denoted by $\Lambda$.

The thermal and velocity boundary conditions now become
\be
\theta_s(z = 1+d_0, t) = - \Lambda \quad \mbox{and} \quad  \quad \theta_s(z = 1, t) = 0; 
\label{eqn:temp_solid}
\ee
\be
\theta_l(z = 0, t) = 1 \quad \mbox{and} \quad \theta_l(z = 1, t) = 0;
\label{eqn:temp_liquid}
\ee
\be
u(z = 0, t) = 1; \, v(z = 0, t) = w(z = 0, t) = 0 \quad \mbox{and} \quad 
\label{eqn:vel_bottom}
\ee
\be
u(z = 1, t) = v(z = 1, t) = w(z = 1, t) = 0. 
\label{eqn:vel_top}
\ee

\section{Linear stability analysis}
We now perform linear stability analysis on equations \ref{eqn:mass_scaled} -- \ref{eqn:stefan_scaled}, with the boundary conditions \ref{eqn:temp_solid} -- \ref{eqn:vel_top}.  

\subsection{Base-state solutions}
All variables in the base state are assumed to be steady and horizontally homogeneous.

\subsubsection{Liquid}
The base-state velocity and temperature profiles are taken to be $u^{(0)}(z)$ and $\theta_l^{(0)}(z)$. Solving the equations of motion subject to the boundary conditions gives
\be
u^{(0)}(z) = 1 - z \quad \mbox{and} \quad \theta_l^{(0)} = 1 - z.
\ee

\subsubsection{Solid}
The solution to the heat equation for the base-state temperature field in the solid is given by
\be
\theta_s^{(0)} = \frac{\Lambda}{d_0} \, \left(1 - z \right).
\ee

\subsubsection{Interface}
In the base state, we assume that the heat fluxes away from and towards the interface balance, so that the initial thickness of the solid layer is constant. Hence, the Stefan condition is
\be
\left[\frac{\mathrm{d} \theta^{(0)}_s}{\mathrm{d}z} - \frac{\mathrm{d} \theta^{(0)}_l}{\mathrm{d}z}\right]_{z = 1} = 0,
\ee
which gives 
\be
d_0 = \Lambda.
\ee

\subsection{Equations for the perturbation amplitudes}
We introduce a normal mode perturbation of the interface given by
\be
h(x,y,t) = 1 + \epsilon \, \exp{\left(i \, k \, x + i \, m \, y + \sigma \, t_1\right)}; \quad \epsilon \ll 1.
\ee
This in turn leads to perturbations in the liquid and solid layers so that the total velocity, pressure, and temperature fields become
\def\A{
\begin{bmatrix}
    u(x,y,z,t) \\
    v(x,y,z,t) \\
    w(x,y,z,t) \\
    p(x,y,z,t) \\
    \theta_l(x,y,z,t) \\
    \theta_s(x,y,z,t) 
\end{bmatrix}}
\def\B{
\begin{bmatrix}
    u^{(0)}(z) \\
    0 \\
    0\\
    p^{(0)}(z) \\
    \theta_l^{(0)}(z) \\
    \theta_s^{(0)}(z) 
\end{bmatrix}}
\def\C{
\begin{bmatrix}
\widehat{u}(z) \\
\widehat{v}(z) \\
\widehat{w}(z) \\
\widehat{p}(z) \\
\widehat{\theta_l}(z) \\
\widehat{\theta_s}(z)
\end{bmatrix}}
\be
\A = \B + \epsilon \, \C \, \exp{\left(i \, k \, x + i \, m \, y + \sigma \, t_1\right)},
\label{eqn:perturb}
\ee

where $t_1 = \frac{1}{\mathcal{S}} \, t$. The range of $\mathcal{S}$ in the experiments of \citet{Gilpin1980} was $\mathcal{S} \approx [4, 7]$, and  hence we are interested in the limit $\mathcal{S} \gg 1$, in which case the rate-controlling process is the release of latent heat, wherein the dynamics in the solid and liquid regions become quasi-steady \citep[e.g.,][]{feltham1999}.

\subsubsection{Liquid}
Linearizing equations \ref{eqn:NS_liquid_scaled} -- \ref{eqn:heat_liquid_scaled} and using equation \ref{eqn:perturb}, we obtain the following equations for the amplitudes:
\be
i \, k \, \widehat{u} + i \, m \, \widehat{v} + D\widehat{w} = 0;
\label{eqn:mass_normal}
\ee
\be
Pe \, \left[i \, k \, u^{(0)} \, \widehat{u} - \widehat{w}\right] = - i \, k \, \widehat{p} + Pr \, \left(D^2 - \gamma^2\right)\widehat{u};
\label{eqn:xmom_normal}
\ee
\be
Pe \, \left[i \, k \, u^{(0)} \, \widehat{v}\right] = - i \, m \, \widehat{p} + Pr \, \left(D^2 - \gamma^2\right)\widehat{v};
\label{eqn:ymom_normal}
\ee
\be
Pe \, \left[i \, k \, u^{(0)} \, \widehat{w}\right] = - D\widehat{p} + Pr \, \left(D^2 - \gamma^2\right)\widehat{w} + \frac{Ra \, Pr}{Pe} \, \widehat{\theta_l};
\label{eqn:zmom_normal}
\ee
\be
Pe \, \left[i \, k \, u^{(0)} \, \widehat{\theta_l} - \widehat{w}\right] = \left(D^2 - \gamma^2\right)\widehat{\theta_l},
\label{eqn:temp_normal}
\ee
where $D \equiv \frac{\mathrm{d}}{\mathrm{d}z}$ and $\gamma^2 = k^2 + m^2$. 
The boundary conditions become
\be
\widehat{u} = \widehat{v} = \widehat{w} = \widehat{\theta_l} = 0 \quad \mbox{at} \quad z = 0,
\label{eqn:boundary_norm_liq_bottom}
\ee
and 
\be
\widehat{u} = 1, \quad \widehat{v} = \widehat{w} = 0; \quad \widehat{\theta_l} = 1 \quad \mbox{at} \quad z = 1.
\label{eqn:boundary_norm_liq_top}
\ee

We now obtain a single equation for $\widehat{\theta_l}$. Following \citet{forth1989}, we eliminate $\widehat{u}$ and $\widehat{v}$  from equations \ref{eqn:xmom_normal} and \ref{eqn:ymom_normal} to obtain
\be
Pe \, \left[i \, k \, u^{(0)} \, \left(-D \widehat{w}\right) - i \, k \, \widehat{w}\right] = \gamma^2 \, \widehat{p} + Pr \, \left(D^2 - \gamma^2 \right) \left(-D\widehat{w}\right).
\label{eqn:eliminate_u_v}
\ee
Eliminating $\widehat{p}$ from equations \ref{eqn:eliminate_u_v} and \ref{eqn:zmom_normal} we obtain
\be
Pe \, \left[i \, k \, u^{(0)} \, \left(D^2 - \gamma^2\right)\widehat{w}\right] = Pr \, \left(D^2 - \gamma^2\right)^2\widehat{w} - \gamma^2 \, \frac{Ra \, Pr}{Pe} \, \widehat{\theta_l}.
\label{eqn:w}
\ee
Finally, eliminating $\widehat{w}$ from equations \ref{eqn:temp_normal} and \ref{eqn:w} gives the following sixth-order ordinary differential equation for $\widehat{\theta_l}$
\begin{widetext}
\be
\begin{split}
0 &= Pr \, D^6 \widehat{\theta_l} - \left[3 \, Pr \, \gamma^2 + i \, k \, u^{(0)} \, Pe \, \left(1 + Pr \right)\right] \, D^4 \widehat{\theta_l} + \left(4 \, i \, k \, Pr \, Pe\right) \, D^3\widehat{\theta_l} \\
&+ \left[3 \, Pr \, \gamma^4 \, + 2 \, i \, k \, \gamma^2 \, u^{(0)} \, Pe \, \left(1 + Pr\right) - k^2 \, Pe^2 \, \left(u^{(0)}\right)^2 \right] \, D^2 \widehat{\theta_l} \\
&- \left(4 \, i \, k \, \gamma^2 Pr \, Pe - 2 \, k^2 \, Pe^2 \, u^{(0)}\right) \, D\widehat{\theta_l}
- \left[Pr \, \gamma^6 + i \, k \, \gamma^4 \, Pe \, u^{(0)} \, \left(1 + Pr\right) - k^2 \, Pe^2 \, \gamma^2 \, \left(u^{(0)}\right)^2 - Ra \, Pr \, \gamma^2 \right] \, \widehat{\theta_l}.
\label{eqn:temp_final}
\end{split}
\ee
\end{widetext}
The boundary conditions at $z = 0$ are
\begin{eqnarray}
\widehat{\theta_l} &=& 0, \label{eqn:boundary1-final} \\
D^2\widehat{\theta_l} &=& 0 \qquad \textrm{and} \\
D^3 \widehat{\theta_l} - \left(i \, k \, Pe + \gamma^2 \right) \, D \widehat{\theta_l} &=&0,
\end{eqnarray}
and those at $z = 1$ are 
\begin{eqnarray}
\widehat{\theta_l} &=& 1, \\
D^2\widehat{\theta_l} - \gamma^2 &=& 0 \qquad \textrm{and} \\
D^3 \widehat{\theta_l} - \gamma^2 \, D \widehat{\theta_l} &=& 0.
\label{eqn:boundary2-final}
\end{eqnarray}

Equation \ref{eqn:temp_final}, along with boundary conditions \ref{eqn:boundary1-final} and \ref{eqn:boundary2-final}, is solved numerically using \emph{Chebfun}  \citep{driscoll2008}.

\subsubsection{Solid}
The equation for $\theta_s'$ is 
\be
\nabla^2 \theta_s' = 0.
\ee
Using normal modes $\theta_s' = \widehat{\theta_s} \, \exp(i \, k \, x + i \, m \, y + \sigma \, t_1)$, we have
\be
\left(D^2 - \gamma^2\right)\widehat{\theta_s} = 0,
\label{eqn:solid-final}
\ee
with
\be
\widehat{\theta_s}\left(z=1\right) = 1 \quad \mbox{and} \quad \widehat{\theta_s}\left(z = 1+d_0\right) = 0
\ee
as the boundary conditions. Equation \ref{eqn:solid-final} has solution
\be
\widehat{\theta_s} = C_1 \, \exp(\gamma \, z) + C_2 \, \exp(-\gamma \, z),
\ee
where
\begin{widetext}
\be
C_1 = - \frac{\exp(-2 \, \gamma \, \left[1 + d_0\right])}{\exp(-\gamma) - \exp(- \gamma - 2 \, \gamma \, d_0)} \quad \mbox{and} \quad C_2 = \frac{1}{\exp(-\gamma) - \exp(- \gamma - 2 \, \gamma \, d_0)}.
\ee
\end{widetext}

\subsubsection{Interface}
At $\mathcal{O}(\epsilon)$ the Stefan condition becomes
\be
\sigma = \frac{1}{\Lambda} \, \left[\frac{\mathrm{d} \widehat{\theta}_s}{\mathrm{d}z} - \frac{\mathrm{d} \widehat{\theta}_l}{\mathrm{d}z}\right]_{z = 1}, 
\label{eqn:stefan-final}
\ee
from which it is evident that the heat flux from the liquid has considerable influence on the stability of the interface. Thus, generation of fluid motions with appreciable vertical velocities can lead to a larger perturbed heat flux, thereby making the interface unstable.

\section{Results and discussion}
\subsection{Phase change with no shear flow}
On setting $Pe = 0$ the present problem reduces to that of phase change in the presence of an unstably stratified column of liquid, which has been studied by \citet{davis1984}. When $Pe = 0$, equation \ref{eqn:temp_final} is independent of $Pr$ and hence so too is the critical Rayleigh number, $Ra_c$, at which convective motions develop \citep{davis1984, chandra2013}. However, as shown by \citet{davis1984}, $Ra_c$ and the critical wavenumber, $\gamma_c$, are functions of $d_0$. 

\begin{figure}
\begin{centering}
\includegraphics[trim = 0 0 0 0, width = \linewidth]{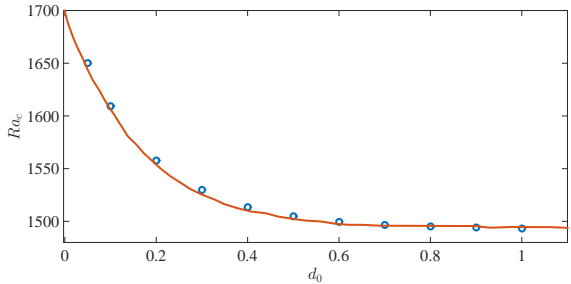} 
\caption{Comparison of $Ra_c(d_0)$ with \citet{davis1984}. Circles are the values from the present calculations, and the solid line is from \citet{davis1984}.}
\label{fig:davis-Ra}
\end{centering}
\end{figure}

\begin{figure}
\begin{centering}
\includegraphics[trim = 0 0 0 0, width = \linewidth]{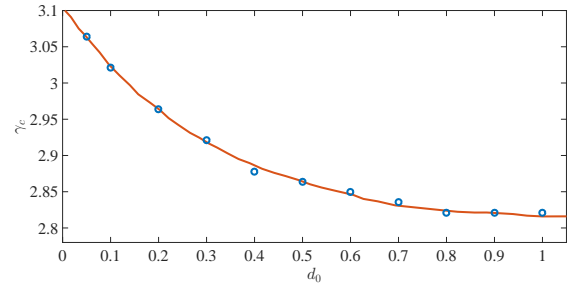} 
\caption{Comparison of $\gamma_c(d_0)$ with \citet{davis1984}. Circles are the values from the present calculations, and the solid line is from \citet{davis1984}.}
\label{fig:davis-k}
\end{centering}
\end{figure}

In Figures \ref{fig:davis-Ra} and \ref{fig:davis-k} we compare $Ra_c$ and $\gamma_c$ as functions of $d_0$ with the calculations of \citet{davis1984}. The decrease in $Ra_c$ with increasing $d_0$ is due to the fact that the velocity boundary condition at the top surface for the liquid is `relaxed' due to the presence of the moving boundary. The calculations of \citet{davis1984} were focused on experiments using cyclohexane, for which we estimate $\mathcal{S} \approx 6 - 8$, showing good agreement with our calculations in figures \ref{fig:davis-Ra} and \ref{fig:davis-k}.

\subsection{Effects of shear}

\subsubsection{Roll structure and its dependence on shear and perturbation wave-vector}

\begin{figure}
\centering
\includegraphics[width = \linewidth]{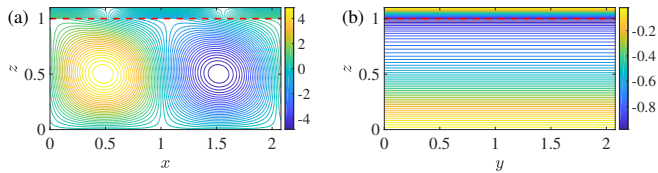} 
\caption{Perturbed temperature field for $Ra = 1700$, $Pe = 0.5$, $\gamma = 3$, and $k = \gamma, \, m = 0$ in (a) $x$-$z$ plane and (b) $y-z$ plane. This case corresponds to $u^{(0)}_ {sq} = u^{(0)}$. The dashed line denotes the solid-liquid interface.}
\label{fig:w1}
\end{figure}

\begin{figure}
\centering
\includegraphics[width = \linewidth]{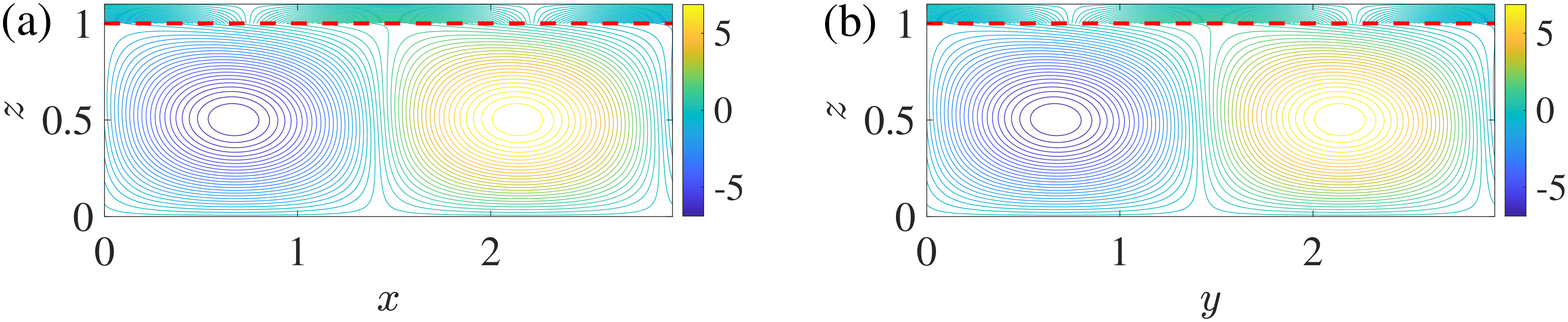} 
\caption{Perturbed temperature field for $Ra = 1700$, $Pe = 0.5$, $\gamma = 3$, and $k = m = \gamma/\sqrt{2}$ in (a) $x$-$z$ plane and (b) $y-z$ plane. This case corresponds to $u^{(0)}_ {sq} < u^{(0)}$. The dashed line denotes the solid-liquid interface.}
\label{fig:w2}
\end{figure}

\begin{figure}
\centering
\includegraphics[width = \linewidth]{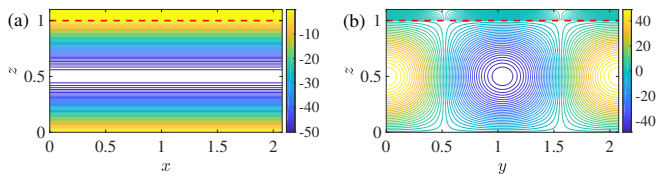} 
\caption{Perturbed temperature field for $Ra = 1700$, $Pe = 0.5$, $\gamma = 3$, and $k = 0, \, m = \gamma$ in (a) $x$-$z$ plane and (b) $y-z$ plane. This case corresponds to $u^{(0)}_ {sq} = 0$. The dashed line denotes the solid-liquid interface.}
\label{fig:w3}
\end{figure}

The effects of shear flow on the perturbations at the interface depend on how the flow is aligned with respect to the perturbation wave-vector $\boldsymbol{\gamma} = (k, m)$ \citep{chung2001, neufeld2008shear}.
This dependence can be understood by following \citet{chung2001} and performing a Squire transformation of the base-state velocity. In our notation this is:
\be
u^{(0)}_ {sq} = \frac{k}{\gamma} \, u^{(0)}.
\ee
Figures \ref{fig:w1}, \ref{fig:w2}, and \ref{fig:w3} show the perturbed temperature field for $Ra = 1700$, $Pe = 0.5$, $\gamma = 3.021$, and $k = \gamma, \, m = 0$ (figure \ref{fig:w1}), $k = m = \gamma/\sqrt{2}$ (figure \ref{fig:w2}), and $k = 0, \, m = \gamma$ (figure \ref{fig:w3}), respectively. It is clearly seen that when $u^{(0)}_ {sq} = u^{(0)}$ $\left(m = 0\right)$ the rolls are aligned such that their axes are perpendicular to the direction of the flow; when $0 < u^{(0)}_ {sq} < u^{(0)}$ $\left(m \neq 0 \right)$ the roll axes are aligned at a certain angle with the shear flow; and when $u^{(0)}_ {sq} = 0$ $\left(m = \gamma\right)$ the roll axes are parallel to the shear flow.

A closer examination of figure \ref{fig:w3} reveals that when $k = 0$, the roll structure is completely unaffected by shear. Hence, shear has no effect on perturbations with wave-vectors perpendicular to it \citep{chung2001, neufeld2008shear}. Noting this dependence on $\boldsymbol{\gamma} = \left(k, m \right)$, we discuss the results in terms of $\gamma$.

\subsubsection{Effects on the instability}

\begin{figure}
\begin{centering}
\includegraphics[width = \linewidth]{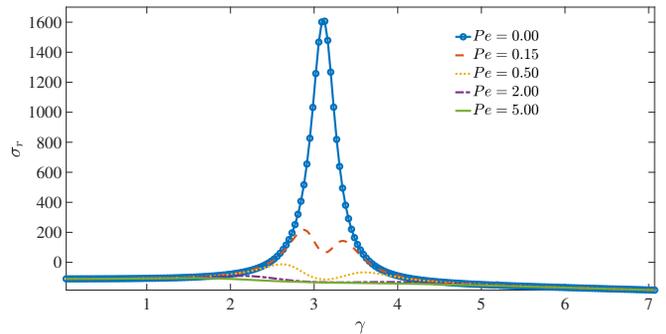} 
\caption{Real growth rates $\sigma_r$ as a function of wavenumber $\gamma$ when $d_0 = 0.1$ and $Ra = 1700$ for different $Pe$.  Shear has a strong stabilizing effect on the instability of the phase boundary. }
\label{fig:sigmar_d01}
\end{centering}
\end{figure}

\begin{figure}
\begin{centering}
\includegraphics[width = \linewidth]{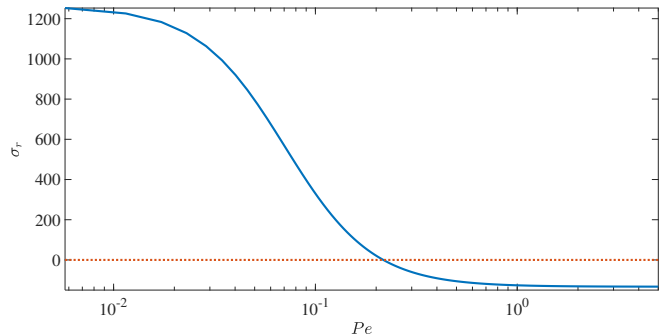} 
\caption{The dependence of $\sigma_r$ on $Pe$ for $d_0 = 0.1$, $Ra = 1700$ and $\gamma = 3.021$, the most rapidly growing mode in the absence of shear flow ($Pe = 0$). The growth rate becomes negative for $Pe \ge 0.22$. (See dotted red line.)}
\label{fig:sigmar_Pe_d01}
\end{centering}
\end{figure}

To understand the effects of shear on the instability of the convective flow, we solve equation \ref{eqn:temp_final} with $Pe = 0, 0.5, 2,$ and $5$ and a supercritical $Ra$ of $1700$. Figure \ref{fig:sigmar_d01} shows the dispersion curve for the real part of the growth rate ($\sigma_r$) for $d_0 = 0.1$ and different $Pe$. In the absence of shear, the most unstable mode has $\gamma = 3.021$, and clearly the interfacial instability is suppressed as the strength of the shear flow increases, with all modes decaying when $Pe$ as small as $0.5$. 
For example, in figure \ref{fig:sigmar_Pe_d01}, 
we see that the growth rate becomes negative for $Pe \approx 0.22$, and asymptotes for $Pe \ge 1$.

\begin{figure}
\begin{centering}
\includegraphics[width = \linewidth]{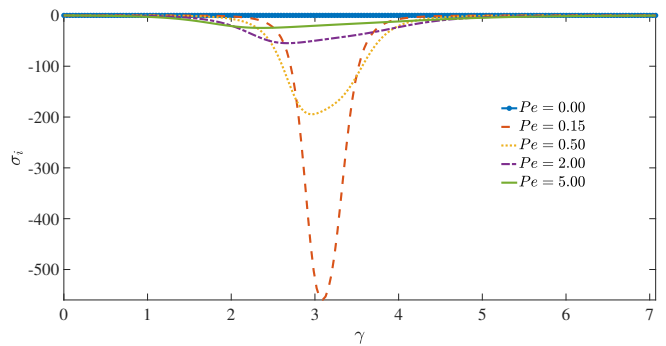} 
\caption{Dispersion curves for $\sigma_i$ with $d_0 = 0.1$ and different $Pe$.}
\label{fig:sigmai_d01}
\end{centering}
\end{figure}

\begin{figure}
\begin{centering}
\includegraphics[width = \linewidth]{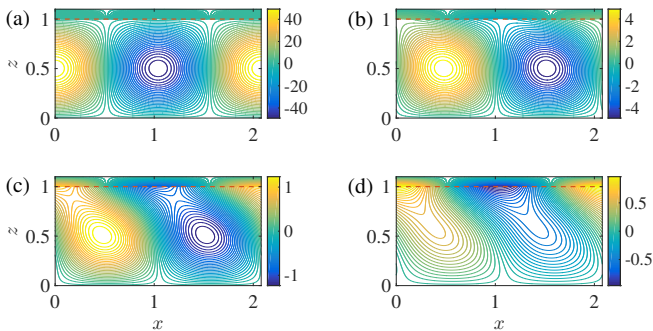} 
\caption{Perturbed temperature field in the liquid and solid regions for $\gamma = 3.021$ $\left(k = m = \gamma/\sqrt{2}\right)$ when $d_0 = 0.1$ and $Ra = 1700$. (a) $Pe = 0.0$, (b) $Pe = 0.5$, $Pe = 2.0$, and (d) $Pe = 5.0$. The dashed lines denote the solid-liquid interface. Qualitatively similar behaviour is also seen for larger values of $d_0$.}
\label{fig:temp_d01}
\end{centering}
\end{figure}

The introduction of shear flow leads to the stabilization of the interface, which is evidenced by the smaller values of $\sigma_r$ relative to those for purely convective flow, and when $\sigma_r > 0$ we find travelling waves along the solid-liquid interface in the direction of the shear flow. As shown in figure \ref{fig:sigmai_d01}, the $\sigma_i(\gamma, Pe)$ curves display non-monotonic behaviour.  This is because in the absence of shear flow, there are no traveling waves and the convective rolls are undistorted.  Thus, for small $Pe$, these rolls are advected by the shear flow with little or no distortion. However, as $Pe$ increases the convective and shear motions interact, leading to the excitation of a larger set of wavenumbers. This causes the convective rolls to lose their structural coherence.
These shear effects can be seen in figure \ref{fig:temp_d01}, which shows the perturbed temperature field in the liquid and solid regions 
as a function of $Pe$.

We should note here that the values of $\sigma$ in the experiments of \citet{Gilpin1980} may have an additional spatial dependence: Because the flow is composed of both shear and buoyancy driven components, the turbulent flow field is spatially inhomogeneous. Hence, a perturbation originating at a particular location at the interface may have the magnitude and/or sign of its growth rate modified as it propagates along the interface. However, a theoretical study of the linear stability of the system avoids this complication.

The results discussed here should be contrasted with those for mushy layers, where stability of the system is a non-monotonic function of the strength of the external shear flow, because of the induced flow within the mushy layer \citep{neufeld2008, neufeld2008shear}. Here, there is no such induced flow and the shear flow only damps perturbations. Namely, the destabilizing factor here is the convective flow that tends to melt the solid phase by enhanced heat transport. The effect of the shear flow is to reduce the strength of vertical motions and hence the upward heat transport. This leads to the decay of perturbations for $Pe \ge 0.22$.

\section{Conclusions}
We have studied the effects of shear and buoyancy driven flow of a pure melt over its solid phase. A reanalysis of the experimental data of \citet{Gilpin1980} shows that the water column in their experiments was unstably stratified, necessitating the inclusion of buoyancy effects to explain the observed phase boundary instability. 
Interpreting the experimental velocity profiles using Monin-Obukhov theory \citep{Monin1971} supports the argument that the effects of buoyancy are important. 

A linear stability analysis of the evolution of buoyancy and shear driven flow of the melt over its solid phase shows that buoyancy is the only destabilizing factor in the system. Shear flow stabilizes the system by reducing the strength of vertical motions and hence vertical heat transport by the convective flow. Our calculations show that for $Pe$ as small as $0.22$, all modes of perturbation decay and the growth rate asymptotes to a negative value for $Pe \ge 1$.  However,  we point out the interesting possibility of a re-entrant interfacial instability at much larger $Pe$, where shear flow instabilities coupled with buoyancy might enhance vertical motions.  

There are clearly implications for situations in which there is a shear flow over a dissolving phase boundary accompanied by a temperature gradient, so that there are potentially three interacting fields (momentum, compositional and thermal) of influence.  Pressure fluctuations associated with interfacial corrugations in mushy layers exposed to shear flow can be relieved by dissolution and solidification of the mushy layer itself \citep{neufeld2008, neufeld2008shear}.  However, when the solid phase is pure, as in the case studied here, imposing a shear flow with impurities and superheat should lead to interesting phenomena since the temperature of maximum density of aqueous solutions depends on impurity concentration.  

Finally, because the $Re$ and (estimated) $Ra$ for their experiments indicate that their flow was in a turbulent regime, our calculations lead us to speculate that the instability observed by \citet{Gilpin1980} may be ostensibly nonlinear; a topic for future study. 

\section*{Acknowledgements}
The authors acknowledge the support of the University of Oxford and Yale University.
S.T. acknowledges a Research Fellowship from All Souls College, Oxford, a NASA Graduate Research Fellowship, and helpful discussions with A. J. Wells.  J.S.W. acknowledges NASA Grant NNH13ZDA001N-CRYO, Swedish Research Council grant no. 638-2013-9243, and a Royal Society Wolfson Research Merit Award for support.

\bibliography{jfm_gilpin}

\begin{thebibliography}{44}%
\makeatletter
\providecommand \@ifxundefined [1]{%
 \@ifx{#1\undefined}
}%
\providecommand \@ifnum [1]{%
 \ifnum #1\expandafter \@firstoftwo
 \else \expandafter \@secondoftwo
 \fi
}%
\providecommand \@ifx [1]{%
 \ifx #1\expandafter \@firstoftwo
 \else \expandafter \@secondoftwo
 \fi
}%
\providecommand \natexlab [1]{#1}%
\providecommand \enquote  [1]{``#1''}%
\providecommand \bibnamefont  [1]{#1}%
\providecommand \bibfnamefont [1]{#1}%
\providecommand \citenamefont [1]{#1}%
\providecommand \href@noop [0]{\@secondoftwo}%
\providecommand \href [0]{\begingroup \@sanitize@url \@href}%
\providecommand \@href[1]{\@@startlink{#1}\@@href}%
\providecommand \@@href[1]{\endgroup#1\@@endlink}%
\providecommand \@sanitize@url [0]{\catcode `\\12\catcode `\$12\catcode
  `\&12\catcode `\#12\catcode `\^12\catcode `\_12\catcode `\%12\relax}%
\providecommand \@@startlink[1]{}%
\providecommand \@@endlink[0]{}%
\providecommand \url  [0]{\begingroup\@sanitize@url \@url }%
\providecommand \@url [1]{\endgroup\@href {#1}{\urlprefix }}%
\providecommand \urlprefix  [0]{URL }%
\providecommand \Eprint [0]{\href }%
\providecommand \doibase [0]{http://dx.doi.org/}%
\providecommand \selectlanguage [0]{\@gobble}%
\providecommand \bibinfo  [0]{\@secondoftwo}%
\providecommand \bibfield  [0]{\@secondoftwo}%
\providecommand \translation [1]{[#1]}%
\providecommand \BibitemOpen [0]{}%
\providecommand \bibitemStop [0]{}%
\providecommand \bibitemNoStop [0]{.\EOS\space}%
\providecommand \EOS [0]{\spacefactor3000\relax}%
\providecommand \BibitemShut  [1]{\csname bibitem#1\endcsname}%
\let\auto@bib@innerbib\@empty
\bibitem [{\citenamefont {Epstein}\ and\ \citenamefont
  {Cheung}(1983)}]{epstein1983}%
  \BibitemOpen
  \bibfield  {author} {\bibinfo {author} {\bibfnamefont {M.}~\bibnamefont
  {Epstein}}\ and\ \bibinfo {author} {\bibfnamefont {F.~B.}\ \bibnamefont
  {Cheung}},\ }\href@noop {} {\bibfield  {journal} {\bibinfo  {journal} {Ann.
  Rev. Fl. Mech.}\ }\textbf {\bibinfo {volume} {15}},\ \bibinfo {pages} {293}
  (\bibinfo {year} {1983})}\BibitemShut {NoStop}%
\bibitem [{\citenamefont {Glicksman}\ \emph {et~al.}(1986)\citenamefont
  {Glicksman}, \citenamefont {Coriell},\ and\ \citenamefont
  {McFadden}}]{glicksman1986}%
  \BibitemOpen
  \bibfield  {author} {\bibinfo {author} {\bibfnamefont {M.~E.}\ \bibnamefont
  {Glicksman}}, \bibinfo {author} {\bibfnamefont {S.~R.}\ \bibnamefont
  {Coriell}}, \ and\ \bibinfo {author} {\bibfnamefont {G.~B.}\ \bibnamefont
  {McFadden}},\ }\href@noop {} {\bibfield  {journal} {\bibinfo  {journal}
  {Annu. Rev. Fl. Mech.}\ }\textbf {\bibinfo {volume} {18}},\ \bibinfo {pages}
  {307} (\bibinfo {year} {1986})}\BibitemShut {NoStop}%
\bibitem [{\citenamefont {Davis}(1990)}]{davis1990}%
  \BibitemOpen
  \bibfield  {author} {\bibinfo {author} {\bibfnamefont {S.~H.}\ \bibnamefont
  {Davis}},\ }\href@noop {} {\bibfield  {journal} {\bibinfo  {journal} {J.
  Fluid Mech.}\ }\textbf {\bibinfo {volume} {212}},\ \bibinfo {pages} {241}
  (\bibinfo {year} {1990})}\BibitemShut {NoStop}%
\bibitem [{\citenamefont {Huppert}(1990)}]{huppert1990}%
  \BibitemOpen
  \bibfield  {author} {\bibinfo {author} {\bibfnamefont {H.~E.}\ \bibnamefont
  {Huppert}},\ }\href@noop {} {\bibfield  {journal} {\bibinfo  {journal} {J.
  Fluid Mech.}\ }\textbf {\bibinfo {volume} {212}},\ \bibinfo {pages} {209}
  (\bibinfo {year} {1990})}\BibitemShut {NoStop}%
\bibitem [{\citenamefont {Worster}(2000)}]{worster2000}%
  \BibitemOpen
  \bibfield  {author} {\bibinfo {author} {\bibfnamefont {M.~G.}\ \bibnamefont
  {Worster}},\ }in\ \href@noop {} {\emph {\bibinfo {booktitle} {Perspectives in
  Fluid Dynamics --- a Collective Introduction to Current Research}}},\
  \bibinfo {editor} {edited by\ \bibinfo {editor} {\bibfnamefont
  {G.}~\bibnamefont {Batchelor}}, \bibinfo {editor} {\bibfnamefont
  {H.}~\bibnamefont {Moffatt}}, \ and\ \bibinfo {editor} {\bibfnamefont
  {M.}~\bibnamefont {Worster}}}\ (\bibinfo  {publisher} {Cambridge University
  Press},\ \bibinfo {year} {2000})\ pp.\ \bibinfo {pages} {393 --
  446}\BibitemShut {NoStop}%
\bibitem [{\citenamefont {Untersteiner}\ and\ \citenamefont
  {Badgley}(1965)}]{Untersteiner:1965}%
  \BibitemOpen
  \bibfield  {author} {\bibinfo {author} {\bibfnamefont {N.}~\bibnamefont
  {Untersteiner}}\ and\ \bibinfo {author} {\bibfnamefont {F.~I.}\ \bibnamefont
  {Badgley}},\ }\href@noop {} {\bibfield  {journal} {\bibinfo  {journal} {J.
  Geophys. Res.}\ }\textbf {\bibinfo {volume} {70}},\ \bibinfo {pages} {4573}
  (\bibinfo {year} {1965})}\BibitemShut {NoStop}%
\bibitem [{\citenamefont {Wettlaufer}(1991)}]{wettlaufer1991}%
  \BibitemOpen
  \bibfield  {author} {\bibinfo {author} {\bibfnamefont {J.~S.}\ \bibnamefont
  {Wettlaufer}},\ }\href@noop {} {\bibfield  {journal} {\bibinfo  {journal} {J.
  Geophys. Res.}\ }\textbf {\bibinfo {volume} {96}},\ \bibinfo {pages} {7215}
  (\bibinfo {year} {1991})}\BibitemShut {NoStop}%
\bibitem [{\citenamefont {McPhee}(2008)}]{mcphee_book}%
  \BibitemOpen
  \bibfield  {author} {\bibinfo {author} {\bibfnamefont {M.~G.}\ \bibnamefont
  {McPhee}},\ }\href@noop {} {\emph {\bibinfo {title} {Air-ice-ocean
  interaction: turbulent ocean boundary layer exchange processes}}}\ (\bibinfo
  {publisher} {Springer},\ \bibinfo {year} {2008})\BibitemShut {NoStop}%
\bibitem [{\citenamefont {Meakin}\ and\ \citenamefont
  {Jamtveit}(2009)}]{meakin2009}%
  \BibitemOpen
  \bibfield  {author} {\bibinfo {author} {\bibfnamefont {P.}~\bibnamefont
  {Meakin}}\ and\ \bibinfo {author} {\bibfnamefont {B.}~\bibnamefont
  {Jamtveit}},\ }\href@noop {} {\bibfield  {journal} {\bibinfo  {journal}
  {Proc. R. Soc. A}\ ,\ \bibinfo {pages} {rspa20090189}} (\bibinfo {year}
  {2009})}\BibitemShut {NoStop}%
\bibitem [{\citenamefont {Solari}\ and\ \citenamefont
  {Parker}(2013)}]{solari2013}%
  \BibitemOpen
  \bibfield  {author} {\bibinfo {author} {\bibfnamefont {L.}~\bibnamefont
  {Solari}}\ and\ \bibinfo {author} {\bibfnamefont {G.}~\bibnamefont
  {Parker}},\ }\href@noop {} {\bibfield  {journal} {\bibinfo  {journal} {J.
  Geophys. Res. Earth Surf.}\ }\textbf {\bibinfo {volume} {118}},\ \bibinfo
  {pages} {1432} (\bibinfo {year} {2013})}\BibitemShut {NoStop}%
\bibitem [{\citenamefont {Ramudu}\ \emph {et~al.}(2016)\citenamefont {Ramudu},
  \citenamefont {Hirsh}, \citenamefont {Olson},\ and\ \citenamefont
  {Gnanadesikan}}]{ramudu2016}%
  \BibitemOpen
  \bibfield  {author} {\bibinfo {author} {\bibfnamefont {E.}~\bibnamefont
  {Ramudu}}, \bibinfo {author} {\bibfnamefont {B.~H.}\ \bibnamefont {Hirsh}},
  \bibinfo {author} {\bibfnamefont {P.}~\bibnamefont {Olson}}, \ and\ \bibinfo
  {author} {\bibfnamefont {A.}~\bibnamefont {Gnanadesikan}},\ }\href@noop {}
  {\bibfield  {journal} {\bibinfo  {journal} {J. Fluid Mech.}\ }\textbf
  {\bibinfo {volume} {798}},\ \bibinfo {pages} {572} (\bibinfo {year}
  {2016})}\BibitemShut {NoStop}%
\bibitem [{\citenamefont {Claudin}\ \emph {et~al.}(2017)\citenamefont
  {Claudin}, \citenamefont {Dur{\'a}n},\ and\ \citenamefont
  {Andreotti}}]{claudin2017}%
  \BibitemOpen
  \bibfield  {author} {\bibinfo {author} {\bibfnamefont {P.}~\bibnamefont
  {Claudin}}, \bibinfo {author} {\bibfnamefont {O.}~\bibnamefont {Dur{\'a}n}},
  \ and\ \bibinfo {author} {\bibfnamefont {B.}~\bibnamefont {Andreotti}},\
  }\href@noop {} {\bibfield  {journal} {\bibinfo  {journal} {J. Fluid Mech.}\
  }\textbf {\bibinfo {volume} {832}} (\bibinfo {year} {2017})}\BibitemShut
  {NoStop}%
\bibitem [{\citenamefont {Delves}(1968)}]{delves1968}%
  \BibitemOpen
  \bibfield  {author} {\bibinfo {author} {\bibfnamefont {R.~T.}\ \bibnamefont
  {Delves}},\ }\href@noop {} {\bibfield  {journal} {\bibinfo  {journal} {J.
  Cryst. Growth}\ }\textbf {\bibinfo {volume} {3}},\ \bibinfo {pages} {562}
  (\bibinfo {year} {1968})}\BibitemShut {NoStop}%
\bibitem [{\citenamefont {Delves}(1971)}]{delves1971}%
  \BibitemOpen
  \bibfield  {author} {\bibinfo {author} {\bibfnamefont {R.~T.}\ \bibnamefont
  {Delves}},\ }\href@noop {} {\bibfield  {journal} {\bibinfo  {journal} {J.
  Cryst. Growth}\ }\textbf {\bibinfo {volume} {8}},\ \bibinfo {pages} {13}
  (\bibinfo {year} {1971})}\BibitemShut {NoStop}%
\bibitem [{\citenamefont {Forth}\ and\ \citenamefont
  {Wheeler}(1989)}]{forth1989}%
  \BibitemOpen
  \bibfield  {author} {\bibinfo {author} {\bibfnamefont {S.~A.}\ \bibnamefont
  {Forth}}\ and\ \bibinfo {author} {\bibfnamefont {A.~A.}\ \bibnamefont
  {Wheeler}},\ }\href@noop {} {\bibfield  {journal} {\bibinfo  {journal} {J.
  Fluid Mech.}\ }\textbf {\bibinfo {volume} {202}},\ \bibinfo {pages} {339}
  (\bibinfo {year} {1989})}\BibitemShut {NoStop}%
\bibitem [{\citenamefont {Davis}\ \emph {et~al.}(1984)\citenamefont {Davis},
  \citenamefont {M{\"u}ller},\ and\ \citenamefont {Dietsche}}]{davis1984}%
  \BibitemOpen
  \bibfield  {author} {\bibinfo {author} {\bibfnamefont {S.~H.}\ \bibnamefont
  {Davis}}, \bibinfo {author} {\bibfnamefont {U.}~\bibnamefont {M{\"u}ller}}, \
  and\ \bibinfo {author} {\bibfnamefont {C.}~\bibnamefont {Dietsche}},\
  }\href@noop {} {\bibfield  {journal} {\bibinfo  {journal} {J. Fluid Mech.}\
  }\textbf {\bibinfo {volume} {144}},\ \bibinfo {pages} {133} (\bibinfo {year}
  {1984})}\BibitemShut {NoStop}%
\bibitem [{\citenamefont {Liu}\ \emph {et~al.}(1996)\citenamefont {Liu},
  \citenamefont {Ning},\ and\ \citenamefont {Ecke}}]{liu1996}%
  \BibitemOpen
  \bibfield  {author} {\bibinfo {author} {\bibfnamefont {Y.}~\bibnamefont
  {Liu}}, \bibinfo {author} {\bibfnamefont {L.}~\bibnamefont {Ning}}, \ and\
  \bibinfo {author} {\bibfnamefont {R.~E.}\ \bibnamefont {Ecke}},\ }\href@noop
  {} {\bibfield  {journal} {\bibinfo  {journal} {Phys. Rev. E}\ }\textbf
  {\bibinfo {volume} {53}},\ \bibinfo {pages} {R5572} (\bibinfo {year}
  {1996})}\BibitemShut {NoStop}%
\bibitem [{\citenamefont {Wettlaufer}\ \emph {et~al.}(1997)\citenamefont
  {Wettlaufer}, \citenamefont {Worster},\ and\ \citenamefont
  {Huppert}}]{wettlaufer1997}%
  \BibitemOpen
  \bibfield  {author} {\bibinfo {author} {\bibfnamefont {J.~S.}\ \bibnamefont
  {Wettlaufer}}, \bibinfo {author} {\bibfnamefont {M.~G.}\ \bibnamefont
  {Worster}}, \ and\ \bibinfo {author} {\bibfnamefont {H.~E.}\ \bibnamefont
  {Huppert}},\ }\href@noop {} {\bibfield  {journal} {\bibinfo  {journal} {J.
  Fluid Mech.}\ }\textbf {\bibinfo {volume} {344}},\ \bibinfo {pages} {291}
  (\bibinfo {year} {1997})}\BibitemShut {NoStop}%
\bibitem [{\citenamefont {Worster}(1997)}]{worster1997}%
  \BibitemOpen
  \bibfield  {author} {\bibinfo {author} {\bibfnamefont {M.~G.}\ \bibnamefont
  {Worster}},\ }\href@noop {} {\bibfield  {journal} {\bibinfo  {journal} {Ann.
  Rev. Fl. Mech.}\ }\textbf {\bibinfo {volume} {29}},\ \bibinfo {pages} {91}
  (\bibinfo {year} {1997})}\BibitemShut {NoStop}%
\bibitem [{\citenamefont {Davies~Wykes}\ \emph {et~al.}(2018)\citenamefont
  {Davies~Wykes}, \citenamefont {Huang}, \citenamefont {Hajjar},\ and\
  \citenamefont {Ristroph}}]{wykes2018}%
  \BibitemOpen
  \bibfield  {author} {\bibinfo {author} {\bibfnamefont {M.~S.}\ \bibnamefont
  {Davies~Wykes}}, \bibinfo {author} {\bibfnamefont {J.~M.}\ \bibnamefont
  {Huang}}, \bibinfo {author} {\bibfnamefont {G.~A.}\ \bibnamefont {Hajjar}}, \
  and\ \bibinfo {author} {\bibfnamefont {L.}~\bibnamefont {Ristroph}},\
  }\href@noop {} {\bibfield  {journal} {\bibinfo  {journal} {Phys. Rev.
  Fluids}\ }\textbf {\bibinfo {volume} {3}},\ \bibinfo {pages} {043801}
  (\bibinfo {year} {2018})}\BibitemShut {NoStop}%
\bibitem [{\citenamefont {Coriell}\ \emph {et~al.}(1984)\citenamefont
  {Coriell}, \citenamefont {McFadden}, \citenamefont {Boisvert},\ and\
  \citenamefont {Sekerka}}]{coriell1984}%
  \BibitemOpen
  \bibfield  {author} {\bibinfo {author} {\bibfnamefont {S.~R.}\ \bibnamefont
  {Coriell}}, \bibinfo {author} {\bibfnamefont {G.~B.}\ \bibnamefont
  {McFadden}}, \bibinfo {author} {\bibfnamefont {R.~F.}\ \bibnamefont
  {Boisvert}}, \ and\ \bibinfo {author} {\bibfnamefont {R.~F.}\ \bibnamefont
  {Sekerka}},\ }\href@noop {} {\bibfield  {journal} {\bibinfo  {journal} {J.
  Cryst. Growth}\ }\textbf {\bibinfo {volume} {69}},\ \bibinfo {pages} {15}
  (\bibinfo {year} {1984})}\BibitemShut {NoStop}%
\bibitem [{\citenamefont {Schulze}\ and\ \citenamefont
  {Davis}(1994)}]{schulze1994}%
  \BibitemOpen
  \bibfield  {author} {\bibinfo {author} {\bibfnamefont {T.~P.}\ \bibnamefont
  {Schulze}}\ and\ \bibinfo {author} {\bibfnamefont {S.~H.}\ \bibnamefont
  {Davis}},\ }\href@noop {} {\bibfield  {journal} {\bibinfo  {journal} {J.
  Cryst. Growth}\ }\textbf {\bibinfo {volume} {143}},\ \bibinfo {pages} {317}
  (\bibinfo {year} {1994})}\BibitemShut {NoStop}%
\bibitem [{\citenamefont {Schulze}\ and\ \citenamefont
  {Davis}(1995)}]{schulze1995}%
  \BibitemOpen
  \bibfield  {author} {\bibinfo {author} {\bibfnamefont {T.~P.}\ \bibnamefont
  {Schulze}}\ and\ \bibinfo {author} {\bibfnamefont {S.~H.}\ \bibnamefont
  {Davis}},\ }\href@noop {} {\bibfield  {journal} {\bibinfo  {journal} {J.
  Cryst. Growth}\ }\textbf {\bibinfo {volume} {149}},\ \bibinfo {pages} {253}
  (\bibinfo {year} {1995})}\BibitemShut {NoStop}%
\bibitem [{\citenamefont {Schulze}\ and\ \citenamefont
  {Davis}(1996)}]{schulze1996}%
  \BibitemOpen
  \bibfield  {author} {\bibinfo {author} {\bibfnamefont {T.~P.}\ \bibnamefont
  {Schulze}}\ and\ \bibinfo {author} {\bibfnamefont {S.~H.}\ \bibnamefont
  {Davis}},\ }\href@noop {} {\bibfield  {journal} {\bibinfo  {journal} {Phys.
  Fluids}\ }\textbf {\bibinfo {volume} {8}},\ \bibinfo {pages} {2319} (\bibinfo
  {year} {1996})}\BibitemShut {NoStop}%
\bibitem [{\citenamefont {Feltham}\ and\ \citenamefont
  {Worster}(1999)}]{feltham1999}%
  \BibitemOpen
  \bibfield  {author} {\bibinfo {author} {\bibfnamefont {D.~L.}\ \bibnamefont
  {Feltham}}\ and\ \bibinfo {author} {\bibfnamefont {M.~G.}\ \bibnamefont
  {Worster}},\ }\href@noop {} {\bibfield  {journal} {\bibinfo  {journal} {J.
  Fluid Mech.}\ }\textbf {\bibinfo {volume} {391}},\ \bibinfo {pages} {337}
  (\bibinfo {year} {1999})}\BibitemShut {NoStop}%
\bibitem [{\citenamefont {Feltham}\ \emph {et~al.}(2002)\citenamefont
  {Feltham}, \citenamefont {Worster},\ and\ \citenamefont
  {Wettlaufer}}]{feltham2002}%
  \BibitemOpen
  \bibfield  {author} {\bibinfo {author} {\bibfnamefont {D.~L.}\ \bibnamefont
  {Feltham}}, \bibinfo {author} {\bibfnamefont {M.~G.}\ \bibnamefont
  {Worster}}, \ and\ \bibinfo {author} {\bibfnamefont {J.~S.}\ \bibnamefont
  {Wettlaufer}},\ }\href@noop {} {\bibfield  {journal} {\bibinfo  {journal} {J.
  Geophys. Res.-Oceans}\ }\textbf {\bibinfo {volume} {107}},\ \bibinfo {pages}
  {1} (\bibinfo {year} {2002})}\BibitemShut {NoStop}%
\bibitem [{\citenamefont {Neufeld}\ \emph {et~al.}(2006)\citenamefont
  {Neufeld}, \citenamefont {Wettlaufer}, \citenamefont {Feltham},\ and\
  \citenamefont {Worster}}]{neufeld2006}%
  \BibitemOpen
  \bibfield  {author} {\bibinfo {author} {\bibfnamefont {J.~A.}\ \bibnamefont
  {Neufeld}}, \bibinfo {author} {\bibfnamefont {J.~S.}\ \bibnamefont
  {Wettlaufer}}, \bibinfo {author} {\bibfnamefont {D.~L.}\ \bibnamefont
  {Feltham}}, \ and\ \bibinfo {author} {\bibfnamefont {M.~G.}\ \bibnamefont
  {Worster}},\ }\href@noop {} {\bibfield  {journal} {\bibinfo  {journal} {J.
  Fluid Mech.}\ }\textbf {\bibinfo {volume} {549}},\ \bibinfo {pages} {442}
  (\bibinfo {year} {2006})}\BibitemShut {NoStop}%
\bibitem [{\citenamefont {Neufeld}\ and\ \citenamefont
  {Wettlaufer}(2008{\natexlab{a}})}]{neufeld2008}%
  \BibitemOpen
  \bibfield  {author} {\bibinfo {author} {\bibfnamefont {J.~A.}\ \bibnamefont
  {Neufeld}}\ and\ \bibinfo {author} {\bibfnamefont {J.~S.}\ \bibnamefont
  {Wettlaufer}},\ }\href@noop {} {\bibfield  {journal} {\bibinfo  {journal} {J.
  Fluid Mech.}\ }\textbf {\bibinfo {volume} {612}},\ \bibinfo {pages} {363}
  (\bibinfo {year} {2008}{\natexlab{a}})}\BibitemShut {NoStop}%
\bibitem [{\citenamefont {Neufeld}\ and\ \citenamefont
  {Wettlaufer}(2008{\natexlab{b}})}]{neufeld2008shear}%
  \BibitemOpen
  \bibfield  {author} {\bibinfo {author} {\bibfnamefont {J.~A.}\ \bibnamefont
  {Neufeld}}\ and\ \bibinfo {author} {\bibfnamefont {J.~S.}\ \bibnamefont
  {Wettlaufer}},\ }\href@noop {} {\bibfield  {journal} {\bibinfo  {journal} {J.
  Fluid Mech.}\ }\textbf {\bibinfo {volume} {612}},\ \bibinfo {pages} {339}
  (\bibinfo {year} {2008}{\natexlab{b}})}\BibitemShut {NoStop}%
\bibitem [{\citenamefont {Camporeale}\ and\ \citenamefont
  {Ridolfi}(2012)}]{camporeale2012}%
  \BibitemOpen
  \bibfield  {author} {\bibinfo {author} {\bibfnamefont {C.}~\bibnamefont
  {Camporeale}}\ and\ \bibinfo {author} {\bibfnamefont {L.}~\bibnamefont
  {Ridolfi}},\ }\href@noop {} {\bibfield  {journal} {\bibinfo  {journal} {J.
  Fluid Mech.}\ }\textbf {\bibinfo {volume} {694}},\ \bibinfo {pages} {225}
  (\bibinfo {year} {2012})}\BibitemShut {NoStop}%
\bibitem [{\citenamefont {Coriell}\ \emph {et~al.}(1980)\citenamefont
  {Coriell}, \citenamefont {Cordes}, \citenamefont {Boettinger},\ and\
  \citenamefont {Sekerka}}]{coriell1980}%
  \BibitemOpen
  \bibfield  {author} {\bibinfo {author} {\bibfnamefont {S.~R.}\ \bibnamefont
  {Coriell}}, \bibinfo {author} {\bibfnamefont {M.~R.}\ \bibnamefont {Cordes}},
  \bibinfo {author} {\bibfnamefont {W.~J.}\ \bibnamefont {Boettinger}}, \ and\
  \bibinfo {author} {\bibfnamefont {R.~F.}\ \bibnamefont {Sekerka}},\
  }\href@noop {} {\bibfield  {journal} {\bibinfo  {journal} {J. Cryst. Growth}\
  }\textbf {\bibinfo {volume} {49}},\ \bibinfo {pages} {13} (\bibinfo {year}
  {1980})}\BibitemShut {NoStop}%
\bibitem [{\citenamefont {Drazin}\ and\ \citenamefont
  {Reid}(2004)}]{drazin2004}%
  \BibitemOpen
  \bibfield  {author} {\bibinfo {author} {\bibfnamefont {P.~G.}\ \bibnamefont
  {Drazin}}\ and\ \bibinfo {author} {\bibfnamefont {W.~H.}\ \bibnamefont
  {Reid}},\ }\href@noop {} {\emph {\bibinfo {title} {Hydrodynamic stability}}}\
  (\bibinfo  {publisher} {Cambridge {U}niversity {P}ress},\ \bibinfo {year}
  {2004})\BibitemShut {NoStop}%
\bibitem [{\citenamefont {Worster}(1991)}]{worster1991}%
  \BibitemOpen
  \bibfield  {author} {\bibinfo {author} {\bibfnamefont {M.~G.}\ \bibnamefont
  {Worster}},\ }\href@noop {} {\bibfield  {journal} {\bibinfo  {journal} {J.
  Fluid Mech.}\ }\textbf {\bibinfo {volume} {224}},\ \bibinfo {pages} {335}
  (\bibinfo {year} {1991})}\BibitemShut {NoStop}%
\bibitem [{\citenamefont {Feltham}\ \emph {et~al.}(2006)\citenamefont
  {Feltham}, \citenamefont {Untersteiner}, \citenamefont {Wettlaufer},\ and\
  \citenamefont {Worster}}]{feltham2006}%
  \BibitemOpen
  \bibfield  {author} {\bibinfo {author} {\bibfnamefont {D.~L.}\ \bibnamefont
  {Feltham}}, \bibinfo {author} {\bibfnamefont {N.}~\bibnamefont
  {Untersteiner}}, \bibinfo {author} {\bibfnamefont {J.~S.}\ \bibnamefont
  {Wettlaufer}}, \ and\ \bibinfo {author} {\bibfnamefont {M.~G.}\ \bibnamefont
  {Worster}},\ }\href@noop {} {\bibfield  {journal} {\bibinfo  {journal}
  {Geophys. Res. Lett.}\ }\textbf {\bibinfo {volume} {33}} (\bibinfo {year}
  {2006})}\BibitemShut {NoStop}%
\bibitem [{\citenamefont {Worster}(1992)}]{worster1992}%
  \BibitemOpen
  \bibfield  {author} {\bibinfo {author} {\bibfnamefont {M.~G.}\ \bibnamefont
  {Worster}},\ }\href@noop {} {\bibfield  {journal} {\bibinfo  {journal} {J.
  Fluid Mech.}\ }\textbf {\bibinfo {volume} {237}},\ \bibinfo {pages} {649}
  (\bibinfo {year} {1992})}\BibitemShut {NoStop}%
\bibitem [{\citenamefont {Gilpin}\ \emph {et~al.}(1980)\citenamefont {Gilpin},
  \citenamefont {Hirata},\ and\ \citenamefont {Cheng}}]{Gilpin1980}%
  \BibitemOpen
  \bibfield  {author} {\bibinfo {author} {\bibfnamefont {R.~R.}\ \bibnamefont
  {Gilpin}}, \bibinfo {author} {\bibfnamefont {T.}~\bibnamefont {Hirata}}, \
  and\ \bibinfo {author} {\bibfnamefont {K.~C.}\ \bibnamefont {Cheng}},\
  }\href@noop {} {\bibfield  {journal} {\bibinfo  {journal} {J. Fluid Mech.}\
  }\textbf {\bibinfo {volume} {99}},\ \bibinfo {pages} {619} (\bibinfo {year}
  {1980})}\BibitemShut {NoStop}%
\bibitem [{\citenamefont {Veronis}(1963)}]{veronis1963}%
  \BibitemOpen
  \bibfield  {author} {\bibinfo {author} {\bibfnamefont {G.}~\bibnamefont
  {Veronis}},\ }\href@noop {} {\bibfield  {journal} {\bibinfo  {journal}
  {Astrophys. J.}\ }\textbf {\bibinfo {volume} {137}},\ \bibinfo {pages} {641}
  (\bibinfo {year} {1963})}\BibitemShut {NoStop}%
\bibitem [{\citenamefont {Toppaladoddi}\ and\ \citenamefont
  {Wettlaufer}(2018)}]{TW2018}%
  \BibitemOpen
  \bibfield  {author} {\bibinfo {author} {\bibfnamefont {S.}~\bibnamefont
  {Toppaladoddi}}\ and\ \bibinfo {author} {\bibfnamefont {J.~S.}\ \bibnamefont
  {Wettlaufer}},\ }\href@noop {} {\bibfield  {journal} {\bibinfo  {journal}
  {Phys. Rev. Fluids}\ }\textbf {\bibinfo {volume} {3}},\ \bibinfo {pages}
  {043501} (\bibinfo {year} {2018})}\BibitemShut {NoStop}%
\bibitem [{\citenamefont {Monin}\ and\ \citenamefont
  {Yaglom}(1971)}]{Monin1971}%
  \BibitemOpen
  \bibfield  {author} {\bibinfo {author} {\bibfnamefont {A.}~\bibnamefont
  {Monin}}\ and\ \bibinfo {author} {\bibfnamefont {A.}~\bibnamefont {Yaglom}},\
  }\href@noop {} {\emph {\bibinfo {title} {Statistical fluid mechanics:
  Mechanics of turbulence volume 1}}}\ (\bibinfo  {publisher} {Dover
  Publications},\ \bibinfo {year} {1971})\BibitemShut {NoStop}%
\bibitem [{\citenamefont {Sreenivasan}(1989)}]{sreenivasan1989}%
  \BibitemOpen
  \bibfield  {author} {\bibinfo {author} {\bibfnamefont {K.~R.}\ \bibnamefont
  {Sreenivasan}},\ }in\ \href@noop {} {\emph {\bibinfo {booktitle} {Frontiers
  in {E}xperimental {F}luid {M}echanics}}}\ (\bibinfo  {publisher} {Springer},\
  \bibinfo {year} {1989})\ pp.\ \bibinfo {pages} {159--209}\BibitemShut
  {NoStop}%
\bibitem [{\citenamefont {Turner}(1979)}]{turner1979}%
  \BibitemOpen
  \bibfield  {author} {\bibinfo {author} {\bibfnamefont {J.~S.}\ \bibnamefont
  {Turner}},\ }\href@noop {} {\emph {\bibinfo {title} {Buoyancy effects in
  fluids}}}\ (\bibinfo  {publisher} {Cambridge {U}niversity {P}ress},\ \bibinfo
  {year} {1979})\BibitemShut {NoStop}%
\bibitem [{\citenamefont {Driscoll}\ \emph {et~al.}(2008)\citenamefont
  {Driscoll}, \citenamefont {Bornemann},\ and\ \citenamefont
  {Trefethen}}]{driscoll2008}%
  \BibitemOpen
  \bibfield  {author} {\bibinfo {author} {\bibfnamefont {T.~A.}\ \bibnamefont
  {Driscoll}}, \bibinfo {author} {\bibfnamefont {F.}~\bibnamefont {Bornemann}},
  \ and\ \bibinfo {author} {\bibfnamefont {L.~N.}\ \bibnamefont {Trefethen}},\
  }\href@noop {} {\bibfield  {journal} {\bibinfo  {journal} {BIT}\ }\textbf
  {\bibinfo {volume} {48}},\ \bibinfo {pages} {701} (\bibinfo {year}
  {2008})}\BibitemShut {NoStop}%
\bibitem [{\citenamefont {Chandrasekhar}(2013)}]{chandra2013}%
  \BibitemOpen
  \bibfield  {author} {\bibinfo {author} {\bibfnamefont {S.}~\bibnamefont
  {Chandrasekhar}},\ }\href@noop {} {\emph {\bibinfo {title} {Hydrodynamic and
  hydromagnetic stability}}}\ (\bibinfo  {publisher} {Dover Publications},\
  \bibinfo {year} {2013})\BibitemShut {NoStop}%
\bibitem [{\citenamefont {Chung}\ and\ \citenamefont {Chen}(2001)}]{chung2001}%
  \BibitemOpen
  \bibfield  {author} {\bibinfo {author} {\bibfnamefont {C.~A.}\ \bibnamefont
  {Chung}}\ and\ \bibinfo {author} {\bibfnamefont {F.}~\bibnamefont {Chen}},\
  }\href@noop {} {\bibfield  {journal} {\bibinfo  {journal} {J. Fluid Mech.}\
  }\textbf {\bibinfo {volume} {436}},\ \bibinfo {pages} {85} (\bibinfo {year}
  {2001})}\BibitemShut {NoStop}%
\end{thebibliography}%

\end{document}